\documentclass[]{aa}

\usepackage{graphicx}
\usepackage{txfonts}
\usepackage{url}
\bibpunct{(}{)}{;}{a}{}{,} 

\newcommand{\Ma}{\ensuremath{M_\mathrm{a}}}
\newcommand{\Mb}{\ensuremath{M_\mathrm{b}}}
\newcommand{\Ra}{\ensuremath{R_\mathrm{a}}}
\newcommand{\Rb}{\ensuremath{R_\mathrm{b}}}
\newcommand{\Ka}{\ensuremath{K_\mathrm{a}}}
\newcommand{\Kb}{\ensuremath{K_\mathrm{b}}}
\newcommand{\Aap}{\ensuremath{a_\mathrm{app}}}
\newcommand{\Aph}{\ensuremath{a_\mathrm{phy}}}

\newcommand{\Msun}{\ensuremath{\mathrm{M_{\sun}}}}

\newcommand{\Rsun}{\ensuremath{\mathrm{R_{\sun}}}}

\newcommand{\Teffa}{\ensuremath{T_\mathrm{a}}}
\newcommand{\Teffb}{\ensuremath{T_\mathrm{b}}}
\newcommand{\Logg}{\ensuremath{\log(g)}}
\newcommand{\Logga}{\ensuremath{\log(g_\mathrm{a})}}
\newcommand{\Loggb}{\ensuremath{\log(g_\mathrm{b})}}

\newcommand{\Kelvin}{\ensuremath{\mathrm{K}}}

\newcommand{\excs}{\extracolsep{\fill}}

\begin{document}

\title{Masses and age of the Chemically Peculiar double-lined binary $\chi$~Lupi
\thanks{Based on data collected with the PIONIER visitor-instrument installed at the ESO Paranal Observatory under programe 088.D-0828}}
\titlerunning{HD141556}

\author{J.-B.~Le~Bouquin\inst{1} \and H. Beust\inst{1} \and G. Duvert\inst{1} \and J.P~Berger\inst{2} \and F. M{\'e}nard\inst{1}\inst{3} \and G.~Zins\inst{1}}

\institute{
  UJF-Grenoble 1 / CNRS-INSU, Institut de Plan{\'e}tologie et d'Astrophysique de Grenoble (IPAG) UMR 5274, Grenoble, France
  \and
  European Organisation for Astronomical Research in the Southern Hemisphere (ESO), Casilla 19001, Santiago 19, Chile
  \and
  CNRS / Univ. de Chile, Laboratoire Franco-Chilien d'Astronomie (LFCA), UMI 3386, Santiago , Chile.
}
\offprints{J.B.~Le~Bouquin\\
  \email{jean-baptiste.lebouquin@obs.ujf-grenoble.fr}}
  
\date{Received; Accepted}

\abstract{}
{We aim at measuring the stellar parameters of the two Chemically Peculiar components of the B9.5Vp HgMn + A2 Vm  double-lined spectroscopic binary HD141556 ($\chi$~Lup), whose period is $15.25$~days.}
{We combined historical radial velocity measurements with new spatially resolved astrometric observations from PIONIER/VLTI to reconstruct the three-dimensional orbit of the binary, and thus obtained the individual masses. We fit the available photometric points together with the flux ratios provided by interferometry to constrain the individual sizes, which we compared to predictions from evolutionary models.}
{The individual masses of the components are $\Ma = 2.84 \pm 0.12\ \Msun$ and $\Mb = 1.94 \pm 0.09\ \Msun$. The dynamical distance is compatible with the Hipparcos parallax. We find linear stellar radii of $\Ra=2.85 \pm 0.15\ \Rsun$ and $\Rb=1.75 \pm 0.18\ \Rsun$. This result validates \emph{a posteriori} the flux ratio used in previous detailed abundance studies. Assuming coevality, we determine a slightly sub-solar initial metallicity $Z=0.012\pm0.003$ and an age of $(2.8\pm0.3)\times10^8\ $years. Finally, our results imply that the primary rotates more slowly than its synchronous velocity, while the secondary is probably synchronous. We show that strong tidal coupling during the pre-main sequence evolution followed by a full decoupling at zero-age main sequence provides a plausible explanation for these very low rotation rates.}
{}


\maketitle


\section{Introduction}

Chemically peculiar stars (CP), usually classified as Bp, Ap, HgMn, or Am stars, have challenged our understanding of stellar atmospheres for decades \citep{preston:1974}. Although not completely understood yet, the origin of the peculiariarities seems closely related to a slow rotation velocity \citep{Abt:1972} and to the presence of a magnetic field that could bring to the atmosphere the stability needed for diffusion processes to be important \citep{Michaud:1970,Michaud:1974}. The CP stars of mercury-manganese (HgMn) type are exceptional cases where no clear magnetic field is detected but peculiarities (and sometimes spots) are still present on the stellar surface. Observations suggest a correlation between the presence of a weak magnetic field, abundance anomalies, and binary properties \citep{Scholler:2010,Hubrig:2012}. More studies of HgMn stars are especially important for understanding the role of the multiplicity and the magnetic field as well as the evolutionary aspects of the chemical peculiarity phenomenon.

The abundance anomalies of the non-magnetic CP star \object{HD141556} ($\chi$~Lup) are among the most extreme of this class. This star provides a unique astrophysical light source for atomic spectroscopy because it is very bright and exhibits numerous very sharp lines in the UV and visible parts of the spectrum. It has been observed at very high signal-to-noise ratio by the Goddard High Resolution Spectrograph on the Hubble Space Telescope and other optical instruments, with the goal to derive abundances that fully span the periodic table of elements. See for instance the paper series by \citet{Wahlgren:1994}, \citet{Leckrone:1999}, \citet{Brandt:1999} \citet{Proffitt:2002}, and \citet{Nielsen:2005} about the so-called $\chi$~Lupi pathfinder project.

Part of the complexity of the task is that HD141556 is in fact a double-lined spectroscopic binary (SB2) composed of a B9.5Vp HgMn primary and an A2 Vm secondary. In agrement with the properties of these classes, the primary seems not to host an organized magnetic field while the secondary does \citep{Mathys:1995}. On the other hand, this characteristic provides the chance to determine dynamically the masses of the individual stars and the distance to the system. With this is mind, this well-studied object becomes a perfect case to compare the stellar parameters with the prediction from evolutionary tracks. However, absolute masses have not been determined so far, most probably because of the rather small size (3\ milliarcseconds) of the apparent orbit. This is the purpose of this paper.

\begin{table*}[t]
\caption{Log of interferometric observations with PIONIER at the VLTI together with the measured astrometric positions of b (faintest) with respect to a (brightest) and flux ratio in the H band. The position angle (PA) is defined north ($0$\ deg) to east ($90$\ deg).}
\label{tab:astrometry}
\centering
\begin{tabular*}{0.98\textwidth}{@{\excs}ccccccccc}
\hline\hline\noalign{\smallskip}
Obs. JD  &       Date & Telescopes     & Spectral &  East &      North &  Separation & PA     &   Flux\\
2400000+  &          &  configuration & setup &  (mas)  &  (mas)    & (mas)          & (deg) &   ratio\\
\noalign{\smallskip}\hline\noalign{\smallskip}
55982.845 & 2012-02-24  & D0-H0-G1-I1 & 3 channels &  $+2.20\pm0.1$ &  $+2.17\pm0.1$ & $3.09\pm0.1$ &$+45.4\pm1.8$& $2.8\pm0.2$ \\
55989.857 & 2012-03-02  & A1-G1-K0-I1 & 7 channels &  $-2.69\pm0.1$ &  $-1.96\pm0.1$  & $3.33\pm0.1$ &$-126.0\pm1.7$& $2.9\pm0.1$ \\
55990.906 & 2012-03-03  & A1-G1-K0-I1 & 3 channels &  $-2.13\pm0.1$ &  $-2.13\pm0.1$  & $3.01\pm0.1$ &$-135.0\pm1.9$& $2.9\pm0.2$ \\
55992.875 & 2012-03-05  & A1-G1-K0-I1 & 3 channels &  $-0.15\pm0.1$ &  $-1.44\pm0.1$  & $1.45\pm0.1$ &$-174.0\pm3.9$& $3.0\pm0.2$\\
\noalign{\smallskip}\hline
\end{tabular*}
\end{table*}

Section~\ref{sec:masses} presents the historically available radial velocity measurements as well as the new astrometric observations obtained with long-baseline interferometry. The data set is used to solve for the dynamical masses of each component and the distance of the system. In Section~\ref{sec:sizes} we constrain the linear radius of each component by fitting the integrated photometry and the new flux ratio in the H band obtained by interferometry. Results are compared with the prediction from evolutionary tracks, providing an estimate of the age and initial metallicity of the system. Section~\ref{sec:circ} discuses the state of orbit circularization and rotation synchronization of the system. The paper ends with brief concluding remarks.

\section{Masses and orbital determination}
\label{sec:masses}

Interferometric data were obtained with the PIONIER combiner \citep[][]{Le-Bouquin:2011} and the four auxiliary telescopes of the Very Large Telescope Interferometer \citep[VLTI,][]{Haguenauer:2010}. Data were dispersed over three or seven spectral channels across the H band (1.50 - 1.80\,$\mu$m). Data were reduced and calibrated with the \texttt{pndrs} package \citep[][]{Le-Bouquin:2011}. Calibration stars were chosen in the JMMC Stellar Diameters Catalog\footnote{\tiny\url{cdsarc.u-strasbg.fr/viz-bin/Cat?II/300}} \citep[JSDC,][]{Lafrasse:2010uq}. A log of the observations can be found in Table~\ref{tab:astrometry}. The amount of observing time per night is about thirty minutes, calibrations included. The calibrated square visibilities and closures phases are shown in Figures~\ref{fig:fit_interfdata_a} to~\ref{fig:fit_interfdata_d}, together with the best-fit binary models. Thanks to the simultaneous combination of four telescopes, the binary is detected in all PIONIER observations with no ambiguity in the relative position of the two components.

We used the \texttt{LITpro}\footnote{\tiny\url{www.jmmc.fr/litpro}} software \citep{Tallon-Bosc:2008} to extract the binary parameters, namely the flux ratio (assumed to be constant across the H band) and the astrometric separation. The expected diameters of the individual components (\mbox{$\approx0.4$\ mas} and \mbox{$\approx0.25$\ mas}) were taken as fixed parameters. This assumption has no significant impact on the results because these diameters are almost unresolved by the longest VLTI baselines.  The uncertainties computed by \texttt{LITpro} are unrealistically small. We performed a bootstrap analysis by randomly selecting subsets of the observations and adding noise to these datasets. The uncertainty is given by the dispersion of the best-fit parameters when fitting these datasets. This procedure was repeated independently for each date. We obtained a typical dispersion of $0.05-0.1$\ mas on the astrometry. Consequently, we assigned a conservative uncertainty of \mbox{$\pm0.1\ $mas} in both right ascension and declination. Results are summarized in Table~\ref{tab:astrometry}.

Table~\ref{tab:rv} summarizes the radial velocity (RV) measurements extracted from \citet{Dworetsky:1972}, which include older data from Campbell and Moore (1928). We arbitrarily assigned uncertainties to the RV measurements by looking at the dispersion of each individual data set in Figure~I from \citet{Dworetsky:1972}.

\begin{table}[t]
\caption{RV data extracted from \citet{Dworetsky:1972}}
\label{tab:rv}
\centering
\begin{tabular*}{0.95\columnwidth}{@{\excs}llll}
\hline\hline\noalign{\smallskip}
Obs. JD & $v_\mathrm{rad,a}$ & $v_\mathrm{rad,b}$ & Error\tablefootmark{a}  \\
2400000+  &    (km\,s$^{-1}$)  &  (km\,s$^{-1}$)  &   (km\,s$^{-1}$)\\
\noalign{\smallskip}\hline\noalign{\smallskip}
18549.50\tablefootmark{b}  &    $+36.6$ &  $-103.3$ &   5.0 \\
18556.52\tablefootmark{b}  &    $-59.5$ &  $+52.0$  &  10.0 \\
19097.88\tablefootmark{b}  &    $+33.2$ &  $-79.1$  &   5.0  \\
19099.87\tablefootmark{b}  &    $+37.7$ &  $-100.1$ &   5.0  \\
19221.71\tablefootmark{b}  &    $+39.2$ &  $-101.9$ &   5.0   \\
37458.809 &   $-51.51$ &  $+43.79$ &   1.0  \\
37832.776 &   $+27.43$ &  $-74.17$ &   1.0  \\
38429.071 &   $+38.34$ &  $-94.19$ &   1.0  \\
38430.043 &   $+37.29$ &  $-94.77$ &   1.0  \\
38447.022 &   $+17.18$ &  $-64.57$ &   1.0  \\
38448.053 &   $ -3.76$ &  $-34.41$ &   3.0  \\
38804.035 &   $-70.25$ &  $+59.17$ &   1.0  \\
33805.032 &   $-60.68$ &  $+44.66$ &   1.5  \\
38838.977 &   $+10.92$ &  $-59.39$ &   1.0  \\
38981.101 &   $+12.35$ &  $-67.50$ &   1.5   \\
\noalign{\smallskip}\hline
\end{tabular*}
\tablefoot{\tablefoottext{a}{Uncertainties were assigned arbitrarily
    based on information available in \citet{Dworetsky:1972}.}
  \tablefoottext{b}{Original RV data from Campbell and
    Moore (1928).}}
\end{table}

\begin{figure*}[!t]
\centering
\includegraphics[width=0.98\textwidth]{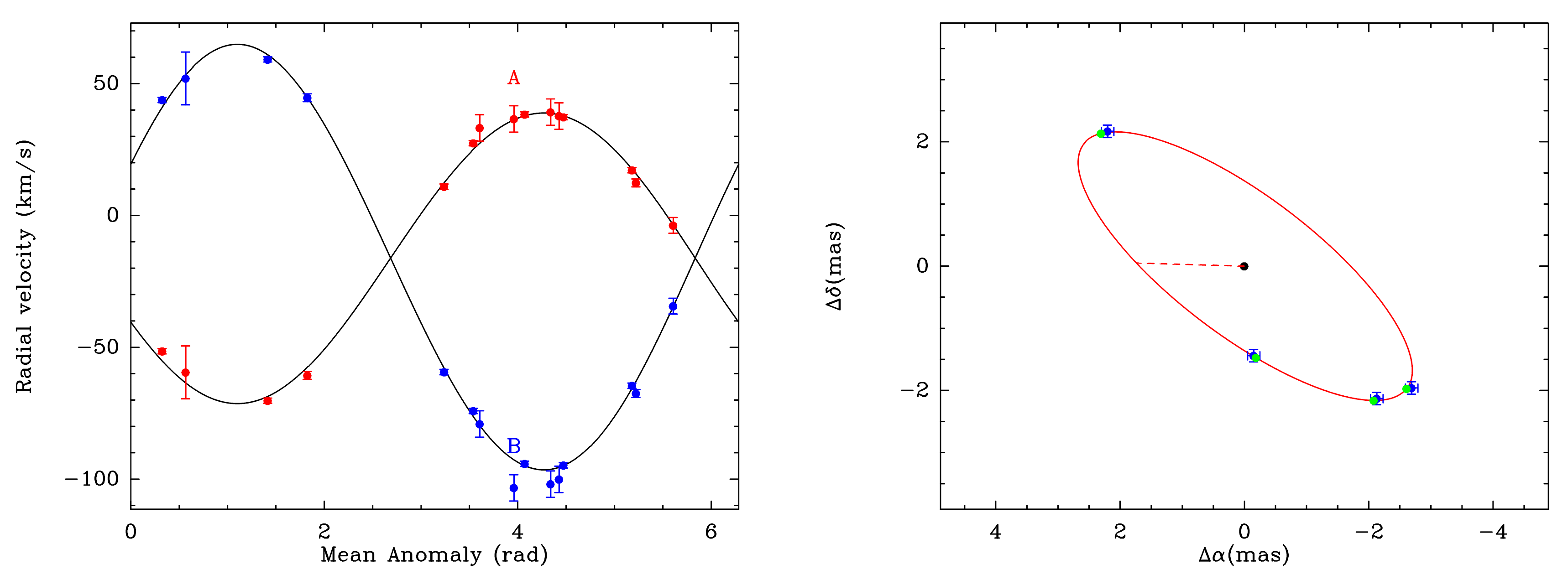}
\caption{Best-fit orbit compared to the historical RV observations (left) and the new 
astrometric observations from PIONIER (right). }\label{fig:fit_orbit}
\end{figure*}

\begin{table}[t]
\caption{Best fit orbital elements and related physical parameters}
\centering
\begin{tabular*}{0.95\columnwidth}{@{\excs}lll}
\hline\hline\noalign{\smallskip}
Parameter & Value  & Unit\\
\noalign{\smallskip}\hline\noalign{\smallskip}
$v_0$ & $-16.06 \pm 0.26$  & km\,s$^{-1}$\\
\Ka &  $55.11\pm 0.46$ & $\mbox{km\,s}^{-1}$ \\
\Kb &  $80.69\pm 0.47$ & $\mbox{km\,s}^{-1}$ \\
$t_\mathrm{p}$ &  $38434.4 \pm 1.7$ & {\tiny JD-2400000}\\
$P$ &  $15.256560 \pm 0.000071$  & days \\
$e$ &  $0.0076 \pm 0.0054$ & \\
$i$ &  $110.2 \pm 2.1$ & $\deg$ \\
$\Omega$ &  $-127.1 \pm 1.5$&$\deg$\\
$\omega$ &  $115.9 \pm 39.7$&$\deg$\\
\Aap &  $3.259 \pm 0.067\:$ & mas \\
Reduced $\chi^2$ & $2.609$\\
\noalign{\smallskip}\hline\noalign{\smallskip}
\Aph &  $0.2030 \pm 0.0029$ & AU \\
\Ma &  $2.84 \pm 0.12$&$\Msun$ \\
\Mb &  $1.94 \pm 0.086$&$\Msun$ \\
$d$ &  $62.27 \pm 1.7\:$&pc \\
\noalign{\smallskip}\hline
\end{tabular*}
\label{tab:bestorbit}
\end{table}
The specificity of the present data set is that we have RV data for the two components of the binary and astrometric data of the relative orbital motion. A combined fit of these data allows one to determine unambiguously all masses and  orbital parameters of the system. In a standard Keplerian model, the radial velocities  $v_\mathrm{rad,a}$ and $v_\mathrm{rad,b}$  of the two stellar components and the $(x,y)$ relative astrometric position (projected onto the sky) of component b relative to component a read
\begin{eqnarray}
v_\mathrm{rad,a} & = & v_0+\Ka\left(\cos(\omega+\nu)+e\cos\omega\right)\;,\\
v_\mathrm{rad,b} & = & v_0-\Kb\left(\cos(\omega+\nu)+e\cos\omega\right)\;,\\
x & = & r_\mathrm{app}\,\left(\cos\Omega\cos(\omega+v)
-\cos i\sin\Omega\sin(\omega+\nu)\right)\;,\\
y & = & r_\mathrm{app}\,\left(\sin\Omega\cos(\omega+v)
+\cos i\cos\Omega\sin(\omega+\nu)\right)\;,\\
\lefteqn{\mbox{with}\qquad r_\mathrm{app}\;=\;\frac{\Aap(1-e^2)}{1+e\cos\nu}.}&&
\end{eqnarray}
Here, \Ka\ and \Kb\ are the individual amplitudes of the radial velocity wobbles of both stars, $v_0$ is the systemic heliocentric offset velocity, $e$ is the orbital eccentricity, and $i$, $\Omega$ and $\omega$ stand for the inclination, longitude of ascending node, and argument of periastron of the orbit with respect to a fixed referential frame. The $OXYZ$ referential frame is chosen in such a way that the $XOY$ plane corresponds to the plane of the sky, that the $OZ$ axis points toward the Earth, and that $\Omega$ is measured counterclockwise from north (the $OX$ axis is $\Delta\delta$ and points toward north; the $OY$ axis is $\Delta\alpha$ and points toward east). \Aap\ is the apparent semi-major axis of the astrometric orbit in milli-arcsec, and $\nu$ is the current true anomaly along the orbit. $\nu$ is related to the time by standard Kepler formalism, via the introduction of the orbital period $P$ (or the orbital angular velocity $n$) and the time of periastron passage $t_p$.

There are therefore ten independent parameters to fit, namely $v_0$, \Ka, \Kb, $t_P$, $n$, $e$, $i$, $\Omega$, $\omega$, and  \Aap. Additional physical quantities can then be derived from the knowledge of these parameters. The RV amplitudes are for instance related to the individual masses $\Ma$ and $\Mb$ by
\begin{equation}
K_\mathrm{a,b}=\frac{M_\mathrm{a,b}}{\Ma+\Mb}\frac{n\ \Aph\sin i}{\sqrt{1-e^2}}\;,
\end{equation}
where \Aph\ is the physical semi-major axis of the orbit (in AU). Thanks to our astrometric data, the inclination $i$ can be fitted as an independent parameter, so that \Aph\ can be simply deduced from $\Ka+\Kb$. Then, with the knowledge of $n$,  the total mass $\Ma+\Mb$ is derived from Kepler's third law, and  the values of \Ka\ and \Kb\ yield the individual masses \Ma\ and \Mb. Finally, the comparison between the physical (\Aph) and apparent (\Aap) semi-major axes gives access to the distance $d$ of the system. 

We performed a combined Levenberg-Marquardt least-square fit of the data. The convergence toward a single and deep $\chi^2$ minimum is rapid and robust for a wide range of initial guesses, so that we are confident in our orbital determination. The error bars on the parameters were deduced from their covariance matrix, which is relevant in the present case  of a deep minimum where the errors are probably Gaussian. The result of this fit is summarized in Fig.~\ref{fig:fit_orbit} and Table~\ref{tab:bestorbit}. 

The very large time separation between the first RV observations (1911) and the last astrometric measurements (2012) allows the orbital period to be determined with a very high accuracy ($\approx 6\:$seconds, i.e., $\approx 5\times10^{-6}$ relative accuracy). The physical separation corresponds to about $15$ stellar radii of the primary. The orbit is compatible with a null eccentricity, because the error bar on the eccentricity is comparable to the fitted value. Consequently, the $\omega$ parameter is meaningless and unconstrained. The estimated distance to the system is compatible with the Hipparcos parallax for this star $16.71\pm0.27 \mathrm{mas} = 59.8\pm1$ pc \citep{van-Leeuwen:2007}. The mass of the primary is slightly larger than the photometric estimate $\Ma=2.72\ \Msun$ from \citet{Guthrie:1986}.

\section{Temperatures, sizes, and age}
\label{sec:sizes}

\begin{figure}
\centering
\includegraphics[scale=0.58]{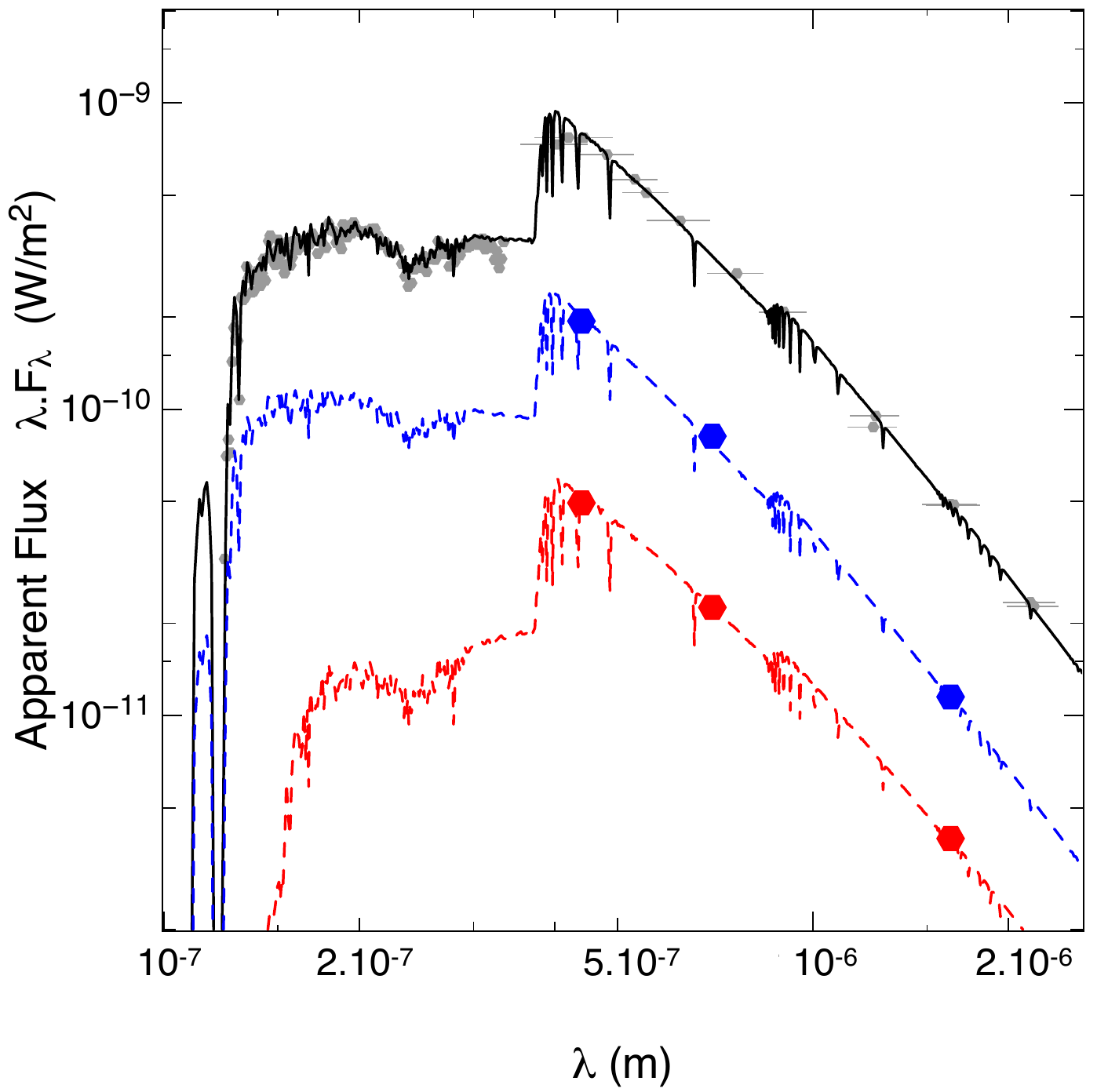}
\caption{Observed SED of the integrated flux of the binary (gray points), modeled by a sum of two Kurucz models with the stellar parameters described in Sect.~\ref{sec:sizes} (solid black line). The red and blue circles are the flux of the primary and the secondary obtained by combining the observed flux ratio with the total flux at nearby wavelengths. These points are lowered by half a decade for clarity purpose. The dashed lines represent the flux of the corresponding Kurucz models. \label{fig:SED} }
\end{figure}

\begin{figure}
\centering
\includegraphics[scale=0.58]{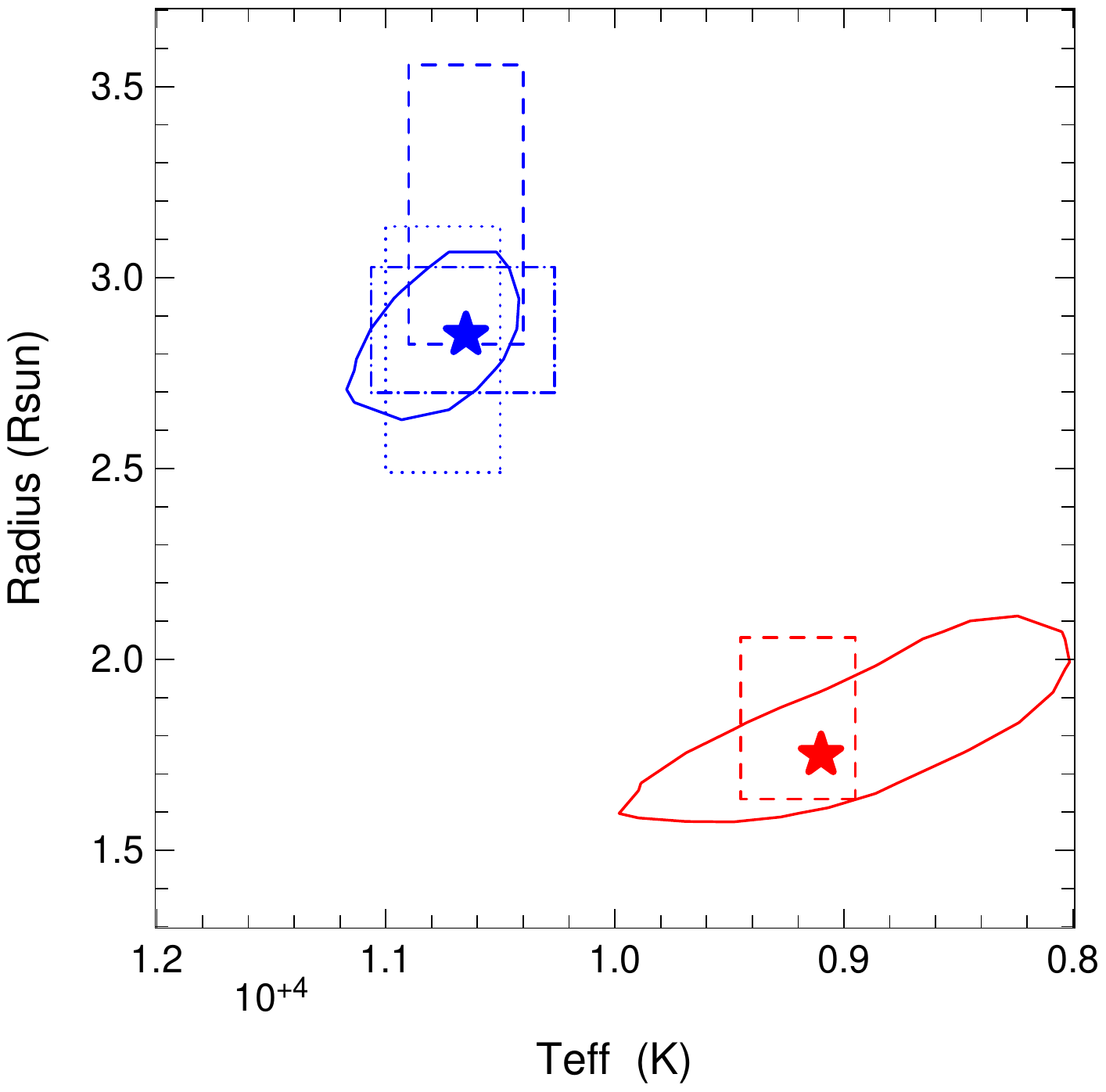}
\caption{Solid lines are the $3\sigma{}$ confidence regions (blue for the primary and red for the secondary) when adjusting the photometry and the flux ratio with Kurucz models. The dot, dashed and dash-dot lines are the confidence region derived from the studies of \citet{Smith:1993}, \citet{Wahlgren:1994}, and \citet{Hubrig:1999} respectively. The filled stars are the adopted stellar parameters.
\label{fig:radius}}
\end{figure}

We retrieved the observed spectral energy distribution (SED) of the sum of the two components using the tool \texttt{VOSpec}\footnote{\tiny \url{www.sciops.esa.int/index.php?project=SAT&page=vospec}} provided by the ESA Virtual Observatory (data coming from the following catalogs: IUE for the UV, Johnson and SDSS for the optical, 2MASS for the near infrared, and IRAS for the infrared). In addition, the interferometric observations presented in the previous section constrain the flux ratio to be $2.9\pm0.1$ in the H band ($1.63\ \mu$m). From spectroscopic analysis in the optical, \citet{Wahlgren:1994} determined a slightly higher flux ratio of $3.92$ at $0.44\ \mu$m and $3.62$ at $0.6\ \mu$m. All these measurements are reported in Fig.~\ref{fig:SED}.

We built synthetic SEDs for each component using the  \texttt{DFSYNTHE}, \texttt{KAPPA9}, \texttt{ATLAS9}, and \texttt{SYNTHE}  codes and tables \citep{Kurucz:1991,sbordone:2004,Castelli:2005}. We used the GNU/LINUX port from Sbordone\footnote{\tiny\url{atmos.obspm.fr}} and Castelli\footnote{\tiny\url{wwwuser.oat.ts.astro.it/castelli}}. For the primary, we furthermore computed a custom opacity density function (ODF) table using the abundances published by \citet{Wahlgren:1994} before running the \texttt{ATLAS9} code. The surface gravity of the models were adjusted to correspond to the masses estimated in Sect.~\ref{sec:masses} although the impact on the result is very small. We used the distance estimated in the previous section to compute the apparent fluxes. Typical values for the interstellar extinction in the solar neighborhood give an extinction of $A(V)=0.05$, which has negligible impact on the results.

The diameters and effective  temperatures of both components were adjusted so that the predictions from the models match the integrated photometric measurement and the available flux ratio. We explored a grid of diameters from one to four solar radii and temperatures from $6\,000\ \Kelvin$ to $15\,000\ \Kelvin$. The result in shown in Fig.~\ref{fig:radius}. The confidence regions of our adjustment are represented by the solid lines. The results for the secondary are partially degenerated because the measurements of the flux ratio lie in the Rayleigh-Jeans domain.

Several estimates of the effective temperatures and the surface gravities for this system exist in the literature, generally using Str\"omgren and Geneva photometry. \citet{Smith:1993} measured the stellar parameters of the primary component and found $\Teffa=10\,750\ \Kelvin$ and $\Logga=4.0$. \citet{Wahlgren:1994} iteratively determined the parameters of the two components by a detailed spectroscopic modeling and obtained $\Teffa=10\,650\ \Kelvin$, $\Teffb=9\,200\ \Kelvin$, $\Logga=3.89$ and $\Loggb=4.2$. \citet{Hubrig:1999} found $\Teffa=10\,664\pm400\ \Kelvin$ and $\Logga=4.22\pm0.05$, when fixing the parameters of the secondary to $\Teffb=9200\ \Kelvin$ and $\Loggb=4.2$ \citep[taken from][]{Wahlgren:1994}. Our independent determination of the masses of each component allows us to convert the surface gravity $g$ and the mass $M$  into the physical radius $R$ of the photosphere:
\begin{equation} 
R = \sqrt{\frac{GM}{g}}\quad,
\end{equation}
where $G$ is the gravitational constant. The resulting confidence region in effective temperature and radii are represented by the dot, dashed, and dashed-dot lines in Fig.~\ref{fig:radius}. When no other information is given in the literature, we assumed an uncertainty of $\pm250\ \Kelvin$ on the temperature and $\pm0.1\:$dex on the $\Logg$. The error boxes overlay adequately with our confidence regions.

To better constrain the stellar radii we decided to adopt the effective temperature of the primary determined from spectroscopy by \citet{Wahlgren:1994}, i.e. $\Teffa=10\,650\ \Kelvin$. Doing so, we obtained a best fit with $\Teffb=9\,100\ \Kelvin$, $\Ra=2.85 \pm 0.15\ \Rsun$ and $\Rb=1.75 \pm 0.18\ \Rsun$. Assuming an uncertainty of $\pm250\ \Kelvin$ on the effective temperature of the primary does not significantly modify the stellar radii. The resulting modeled SED is displayed in Fig.~\ref{fig:SED}, together with the available measurements. Our revision of the stellar radii lead to a ratio $\Ra/\Rb=1.63\pm0.08$. This is in reasonable agreement with the value $\Ra/\Rb=1.67$ derived by \citet{Wahlgren:1994} from the spectroscopic flux ratio. The flux ratio in the B, R, and H bands are all correctly reproduced. Since the latter is a direct and fully independent measurement, it validates \emph{a posteriori} the flux ratio used in the abundance studies from \citet{Leckrone:1999}, \citet{Brandt:1999}, and \citet{Nielsen:2005}.

\begin{figure}[t]
\centering
\includegraphics[scale=0.58]{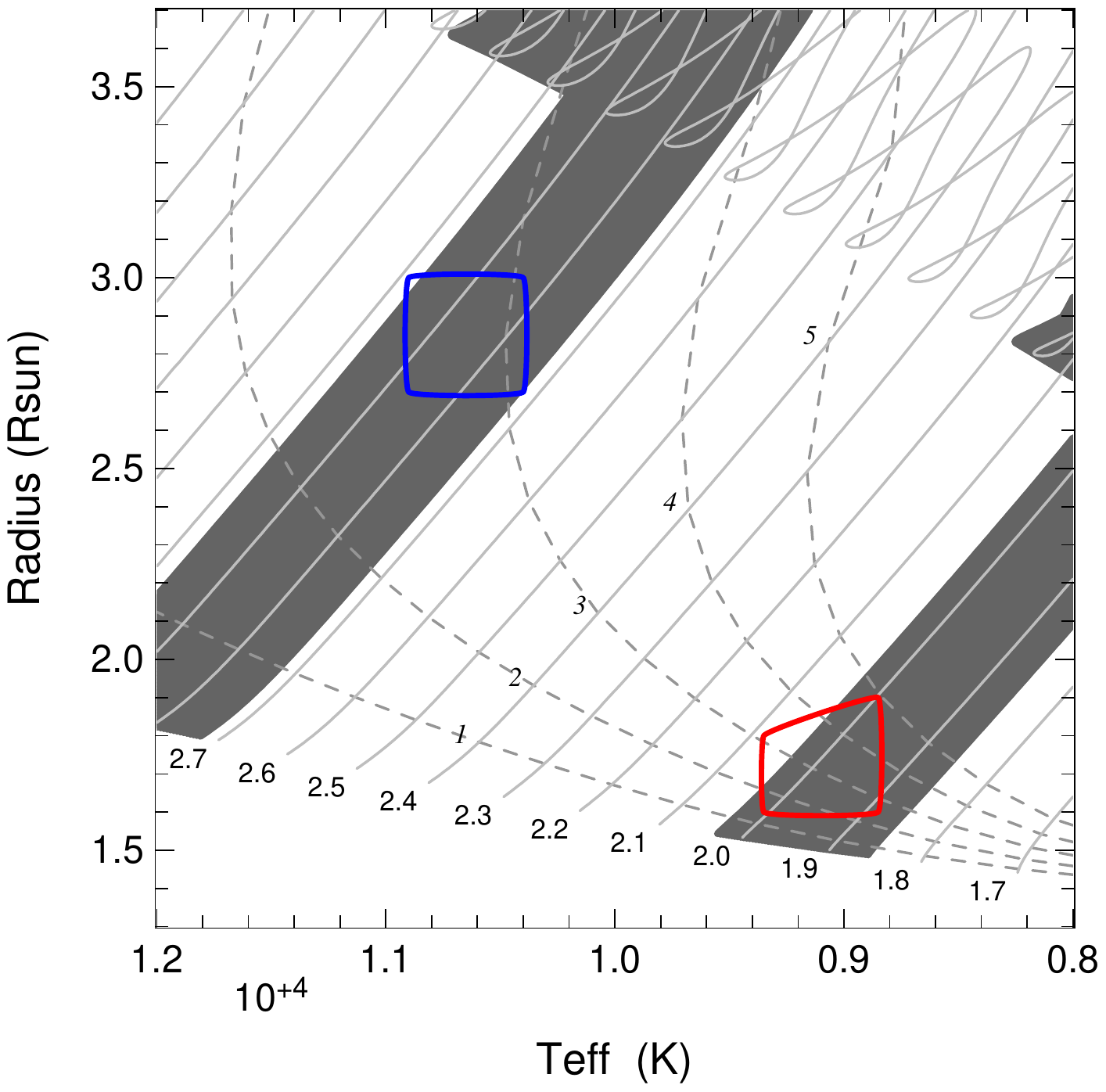}
\caption{Evolutionary tracks from \cite{Mowlavi:2012} for slightly sub-solar initial metallicity ($Z=0.012$). Dashed lines are isochrones for ages from $1$ to $5\times10^8$ years. The gray regions correspond to tracks compatible with the dynamical masses estimated in Sect.~\ref{sec:masses}.  The colored boxes are the effective temperatures and stellar radii discussed in Sect.~\ref{sec:sizes}. } \label{fig:HR}
\end{figure}

\begin{figure}[t]
\centering
\includegraphics[scale=0.58]{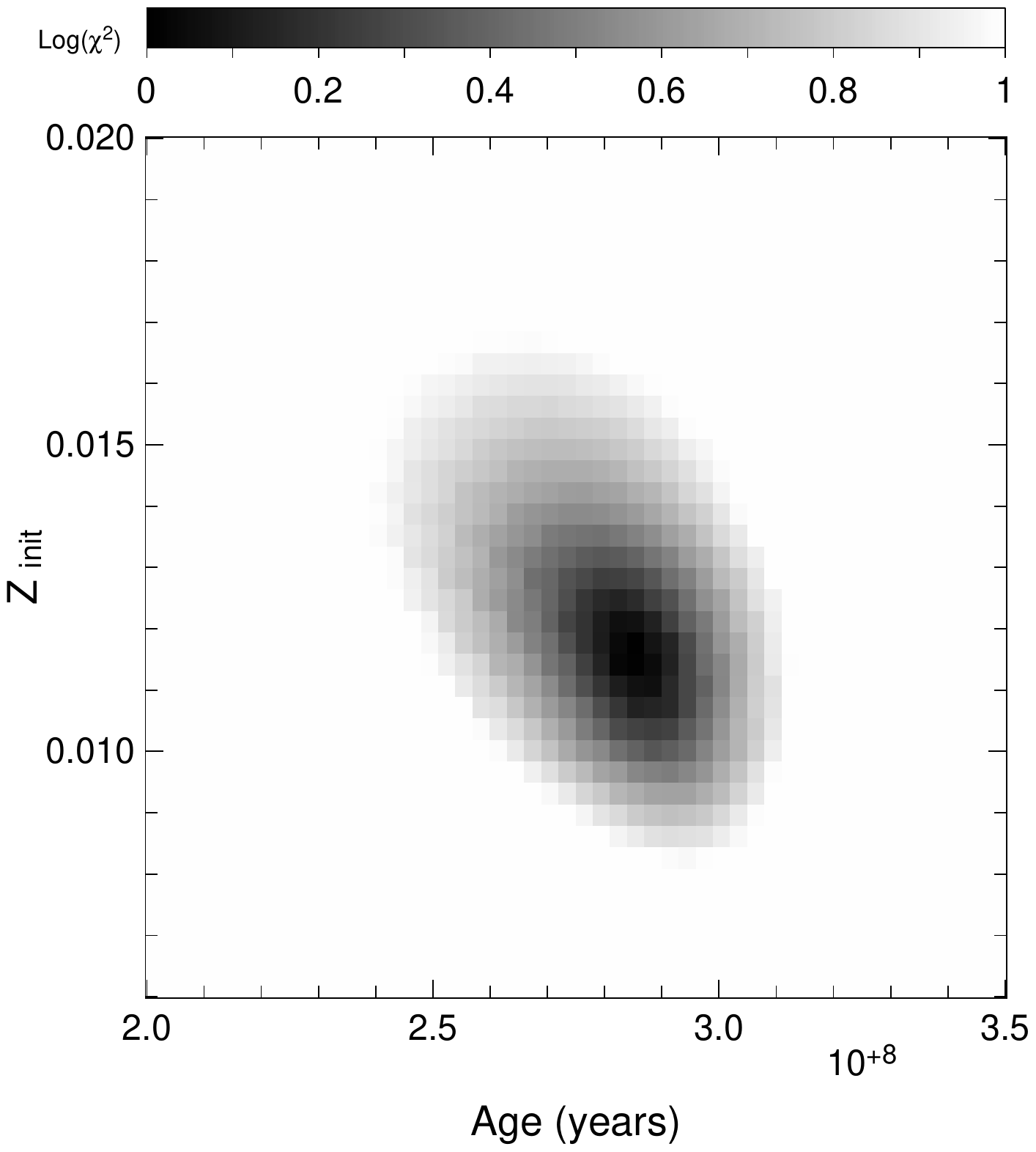}
\caption{Map of $\mathrm{Log}(\chi^2)$\ versus initial metalicity (vertical axis) and age of the system (horizontal axis) when comparing the effective temperatures and radii to the prediction of the evolutionary tracks from \cite{Mowlavi:2012}. We used the dynamical masses estimated in Sect.~\ref{sec:masses} and assumed coevality and identical initial metallicity for the two components.} \label{fig:AgeZ}
\end{figure}

\label{sec:age}
We then compared our estimates of the effective temperatures and stellar radii with predictions from evolutionary tracks computed for the masses determined in Sect.~\ref{sec:masses}. This is shown in Fig.~\ref{fig:HR}. The confidence boxes represent the intersection of the different constraints from Fig.~\ref{fig:radius}. We used the tracks from \cite{Mowlavi:2012} and a slightly sub-solar initial metallicity $Z=0.012$. For both components, the confidence boxes overlap correctly with the evolutionary tracks. We explored how sensitive the agreement is to the chosen initial metallicity. We performed a grid of \mbox{\{Age, $Z$\}}, and compared the predicted radii and effective temperature to the observations. Doing so, we assumed that the two components are coeval and have the same initial metallicity. These are realistic assumptions for such a tight binary of massive stars, whose probability to result from a capture is extremely low. The resulting $\chi^2$ map is shown in Fig.~\ref{fig:AgeZ}. Every initial metallicity in the range $Z=0.009$ to $0.014$ provides a correct agreement between the tracks and the boxes, and respects coevality. The best agreement is formally found for an initial metallicity $Z=0.0115$ and an age of $2.85\times10^8\ $years.

\section{Circularization and synchronization}
\label{sec:circ}
\begin{table}
\caption[]{Angular rotation velocities, assuming the
  radii determined in the present work and $v_\mathrm\sin i_r$ values from the literature.}
\label{tab:rotation}
\begin{tabular*}{\columnwidth}{@{\excs}lllll}
\hline\hline\noalign{\smallskip}
Ref. & $v_\mathrm{a}\sin i_r$ & $v_\mathrm{b}\sin i_r$ &
$N_\mathrm{a}$ & $N_\mathrm{b}$\\
& (km\,s$^{-1}$) & (km\,s$^{-1}$) & ($\mu$rad\,s$^{-1}$) & ($\mu$rad\,s$^{-1}$)\\
\noalign{\smallskip}\hline\noalign{\smallskip}
(1) & $1$    & $2$ & $0.5$ & $1.7$ \\
(2) & $2.5$ & $2$ & $1.4$ & $1.7$\ \\
(3) & $2$    & $4$ & $1.1$ & $3.5$ \\
Average & $1.8\pm1$ & $2.7\pm2$ & $1.0\pm0.6$ & $2.4\pm1.8$\\
\noalign{\smallskip}\hline
\end{tabular*}
\tablebib{(1)~\citet{Leckrone:1999} ; (2)~\citet{Hubrig:1999} ; (3)~\citet{Nielsen:2005}}
\end{table}
The mechanisms responsible for the development of the chemical anomalies in the stellar atmosphere of HgMn stars are not yet fully understood.  Slow rotational velocities are definitely the most important ingredients. For HD141556, the existing studies all found low small rotational velocities but disagree on the exact values (see Table~\ref{tab:rotation}). Measuring such a small broadening is indeed challenging. However, we note that the $v\sin i_r$ are still higher than the microturbulent velocities \citep[$\xi_a=0\ \mbox{km\ s}^{-1}$ and $\xi_b=2\ \mbox{km\ s}^{-1}$,][]{Wahlgren:1994}. This makes us confident that the measured rotational velocities are, at least, reliable upper limits. The literature provides no uncertainties for these quantities and we used instead the dispersion of the existing measurements.

It is worth questioning, then, whether the system has reached such low rotational velocities by means of synchronisation between rotation and orbital motion, thanks to past or still present tidal interaction. To do this, we need to compare the true angular rotation velocities of the stars with the orbital angular velocity $n=4.7\ \mu\mbox{rad\,s}^{-1}$.

Tides in a binary system usually act to align and synchronize spins with orbital motion and at circularizing the orbit. It is well known that spin alignement and synchronization are usually achieved before circularization. The virtually zero eccentricity of the pair is an indication for an already achieved tidal circularization, and thus spin alignement and synchronization. Consequently, we may reasonably assume that the stars have rotation axes perpendicular to their orbital plane (zero obliquities, $i_r = i$). With this assumption and our knowledge of the stellar radii, it is then possible to derive the angular rotation velocities $N_\mathrm{a}$ and $N_\mathrm{b}$ of the stars:
\begin{equation}
  N = \frac{v\sin i_r}{R \sin i}\;\;.
\end{equation}
The results are displayed in Table~\ref{tab:rotation}. An exact synchronism would imply  $N_\mathrm{a}=N_\mathrm{b}=n$. While the secondary could appear to be synchronized within the error bars, obviously the primary rotates sub-synchronously with respect to the orbital motion. Sub-synchronism is rare among short-period binaries, but some cases have already been reported \citep{Casey:1995,Fekel:2011}. Strikingly, the occurence of sub-synchronism seems to be higher in binaries with an HgMn component \citep{Guthrie:1986,Hubrig:1995}. Hence it is interesting to study the history of synchronization in the HD141556 system.

There are different theories for tides in close binary stars that provide predictions for synchronization and circularization times. The first one is the \emph{equilibrium tide}, where each component, distorted by the other, reaches an equilibrium figure. Then the viscosity prevents the equilibrium bulges from permanently pointing toward each other. The result is a coupling between rotation and orbital motion. This theory was first described in detail by \citet{Alexander:1973}. The synchronization and circularization times depend on the mean viscosity of the stars, which are poorly constrained parameters. But \citet{Press:1975} showed that in radiative envelopes, turbulent viscosity dominates, and the authors deduced explicit formulas.

\citet{Zahn:1975} introduced another mechanism called \emph{dynamical tides} that mainly applies to stars with radiative envelopes. In these stars, non-adiabatic oscillations driven by one component onto the other are damped by radiative dissipation, which causes a torque that is responsible for coupling between rotation and orbital motion. \citet{Zahn:1977} gives characteristic times.

A third mechanism was introduced by \citet{Tassoul:1992} and is based on \emph{tidally driven meridional currents within the stars}. According to these authors, this purely hydrodynamical mechanism can naturally explain the high degree of synchronism and circularization observed among close early-type and late-type binaries. They provide characteristic times for synchronization and circularization that mainly depend on the masses, sizes, luminosity, and separation of the components.

\begin{table}[tbp]
\caption[]{Characteristic times for the different tidal 
mechanisms, computed at ZAMS. Details can be found in the appendix.}
\label{times}
\begin{tabular*}{\columnwidth}{@{\excs}llll}
\hline\hline\noalign{\smallskip}
Mechanism (ref.) & $t_{\rm circ}$ & $\left(t_{\rm sync}\right)_\mathrm{a}$ &
$\left(t_{\rm sync}\right)_\mathrm{b}$ \\
& (yrs) & (yrs) & (yrs)\\
\noalign{\smallskip}\hline\noalign{\smallskip}
Equilibrium tide (1) &  $1.37\times10^{17}$ & $3.67\times10^{13}$ & $2.70\times10^{13}$\\ 
Dynamical tide (2) & $5.86\times10^{16}$ & $1.23\times10^{12}$ & $5.04\times10^{11}$\\
Meridional circulation (3) & $1.94\times10^{10}$ & $3.20\times10^6$ &  $1.79\times10^4$\\
\noalign{\smallskip}\hline
\end{tabular*}
\tablebib{(1)~\citet{Press:1975} ; (2)~\citet{Zahn:1977} ; (3)~\citet{Tassoul:1992}}
\end{table}

The characteristic times are summarized in Table~\ref{times}, while the details of the computation can be found in the appendix. The difficulty is always to estimate the various characteristic constants that appear in these formulas. Note that the actual circularization time is computed from the two separate circularization times
\begin{equation}
  \frac{1}{t_\mathrm{circ}}=\frac{1}{\left(t_\mathrm{circ}\right)_\mathrm{a}}+
  \frac{1}{\left(t_\mathrm{circ}\right)_\mathrm{b}}\qquad,
\end{equation}
because the contributions of both stars to the circularization process are added.

The mechanism proposed by \citet{Tassoul:1992} is by far the fastest, as already noted by \citet{Beust:1997}. However, the reality of this mechanism is questioned \citep{Rieutord:1997}. We also note that dynamical tides are more efficient than equilibrium tides. This agrees with \citet{Zahn:2008}, who stressed that dynamical tides are the dominant mechanism in stars with an outer radiative envelope (which is the case here), and that equilibrium tides are stronger for stars with a convective envelope and for planets.

What conclusion can we derive for the HD\,141556 system? First, we note that irrespective of the mechanism, circularization is not expected to be achieved at the age of the system. This could appear to be surprising given the eccentricity we measure, but it is worth recalling that the strength of the tides is highly time-variable. Basically, for such stars, tides are expected to be much stronger in the pre-main sequence (PMS) phase, thanks to the presence of convective envelopes. Circularization is therefore very likely to have enough time to be reached during the PMS phase. But in that case, the stars should also meanwhile have synchronized their rotations. Second, we stress that if the \citet{Tassoul:1992} mechanism were active today, both stars would still be perfectly synchronized. The synchronization times associated to this mechanism are indeed very short. Hence any departure from synchronism should be quickly damped by meridional circulation. Thus the observed sub-synchronism of the primary (whatever its origin) indicates that meridional circulation is very likely inactive in this pair.

Our proposed scenario is then the following: in the PMS phase, the tidal coupling of these stars was strong. It could be due to both dynamical and equilibrium tides. This strong coupling resulted in a rapid synchronization and circularization of the pair. Then, after the zero-age main-sequence (ZAMS) and the settling of a radiative envelope, the tidal coupling dropped drastically. In the absence of any external perturbation to the orbit, the eccentricity remained low.  But between ZAMS and today, both stars underwent a significant radius increase. With our knowledge of the current radii and of the radii at ZAMS (Fig.~\ref{fig:HR}), we can compute the expected angular velocities for the present day $N_\mathrm{now}=n\ (R_\mathrm{zams} / R_\mathrm{now})^2$. This formula assumes perfect synchronism at the ZAMS, conservation of the rotational angular momenta between ZAMS and now, and no change in the internal structures of the stars. We found $N_\mathrm{a}=1.9\,\mu\mbox{rad\,s}^{-1}$ and $N_\mathrm{b}=3.5\,\mu\mbox{rad\,s}^{-1}$. These values are still slightly higher than that of Table~\ref{tab:rotation}, but, at first order, this simplistic scenario provides a plausible explanation for the sub-synchronism of the two components.

We stress that there is no definite theory for the explanation of sub-synchronous rotation. A classical scenario for slowing down stars is of course a magnetic field. The existence of an efficient magnetic braking has been proposed to explain why all the magnetic A stars (Ap) have very slow rotational velocities. In binary system, such a magnetic braking could overcome the tidal synchronisation and naturally lead to sub-synchronous rotation \citep{Hubrig:1995}. This is especially true for HD141556 because the tidal coupling seems weak according to the calculation presented above. However, this scenario requires that the two components are, or were at some point, strongly magnetized. Moreover, it does not explain the almost perfect circularization of the orbit.

\section{Concluding remarks}
\label{sec:conclusions}

HD141556 is a promising target for a precision test of evolutionary models with chemically peculiar stars. Accurate radial velocity measurements (according to modern standards) and a few additional interferometric observations spread over the orbital cycle should considerably reduce the mass uncertainties, which are currently at the $\approx5\%$ level. A single interferometric observation in the visible could provide a direct estimate of the flux ratio in this band (existing estimates rely on spectroscopic analysis), which should help in constraining the temperatures and the ratio of radii. A direct measurement of the stellar diameters (\mbox{0.43\ mas}) is within reach with the SUSI interferometer or a next-generation visible instrument at VLTI.

Our proposed scenario to explain the current rotation velocities implies that the stars have been synchronized by tidal coupling in the final stage of contracting toward the zero-age main sequence, and then fully decoupled during their evolution on the main sequence. Adding generality to this result, it supports the fact that many HgMn stars are found in young multiple systems and in young associations. It is conceivable that these stars have been synchronized and that their peculiarities were established as early as the pre-main sequence phase. A quantitative test of this scenario requires modeling the internal structure of the two stars at all times, including the pre-main sequence.

\begin{acknowledgements} 
PIONIER is funded by the Universit\'e Joseph Fourier (UJF, Grenoble) through its Poles TUNES and SMING and the vice-president of research, the Institut de Plan\'etologie et d'Astrophysique de Grenoble, the ``Agence Nationale pour la Recherche'' with the program ANR EXOZODI, and the Institut National des Science de l'Univers (INSU) with the programs ``Programme National de Physique Stellaire'' and ``Programme National de Plan\'etologie''. The integrated optics beam combiner is the result of a collaboration between IPAG and CEA-LETI based on CNES R\&T funding. FMe acknowledges support from EU FP7-2011 under grant agreement No. 284405 (DIANA) and from the Milenium Nucleus P10-022-F, funded by the Chilean Government. The authors want to warmly thank all the people involved in the VLTI project. It made use of the Smithsonian/NASA Astrophysics Data System (ADS), of the Centre de Donnees astronomiques de Strasbourg (CDS) and of the Jean-Marie Mariotti Center (JMMC). Some calculations and graphics were performed with the freeware \texttt{Yorick}.
\end{acknowledgements}



\appendix

\section{Interferometric observations}

\begin{figure*}
\centering
\includegraphics[width=0.83\textwidth]{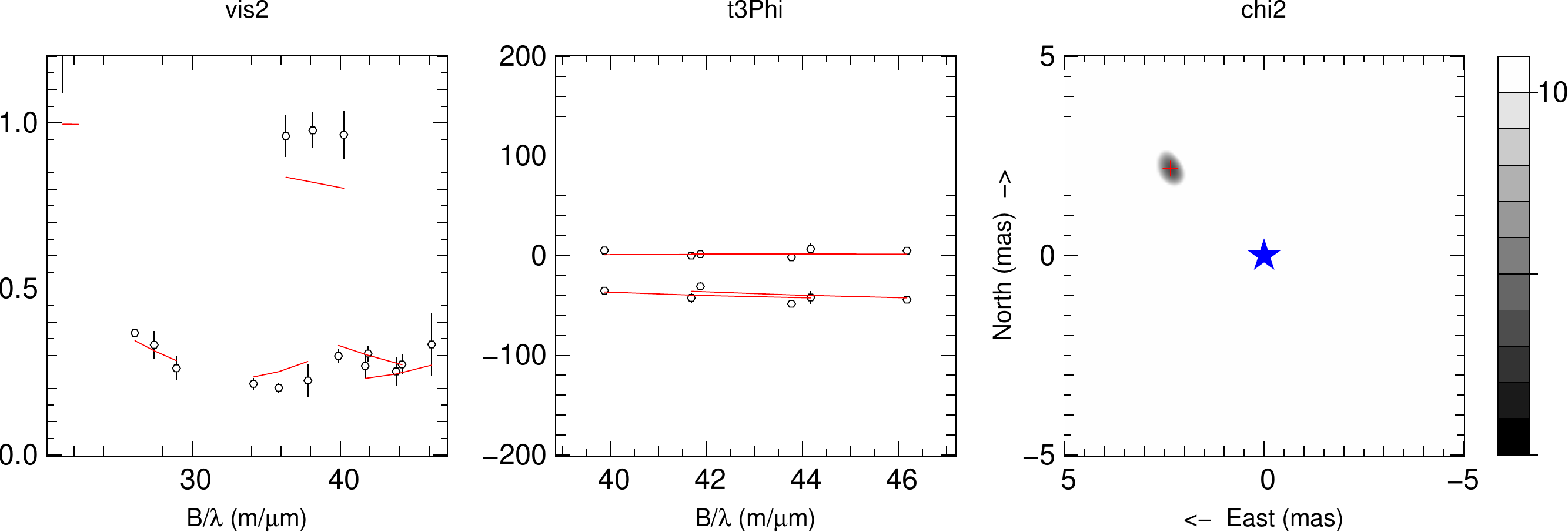}
\caption{Square visibilities (left), closure phases (middle) and corresponding reduced $\chi^2$ map (right) for the PIONIER observation of 2012-02-24. The corresponding best-fit binary model is shown in red.}\label{fig:fit_interfdata_a}\vspace{0.5cm}
\includegraphics[width=0.83\textwidth]{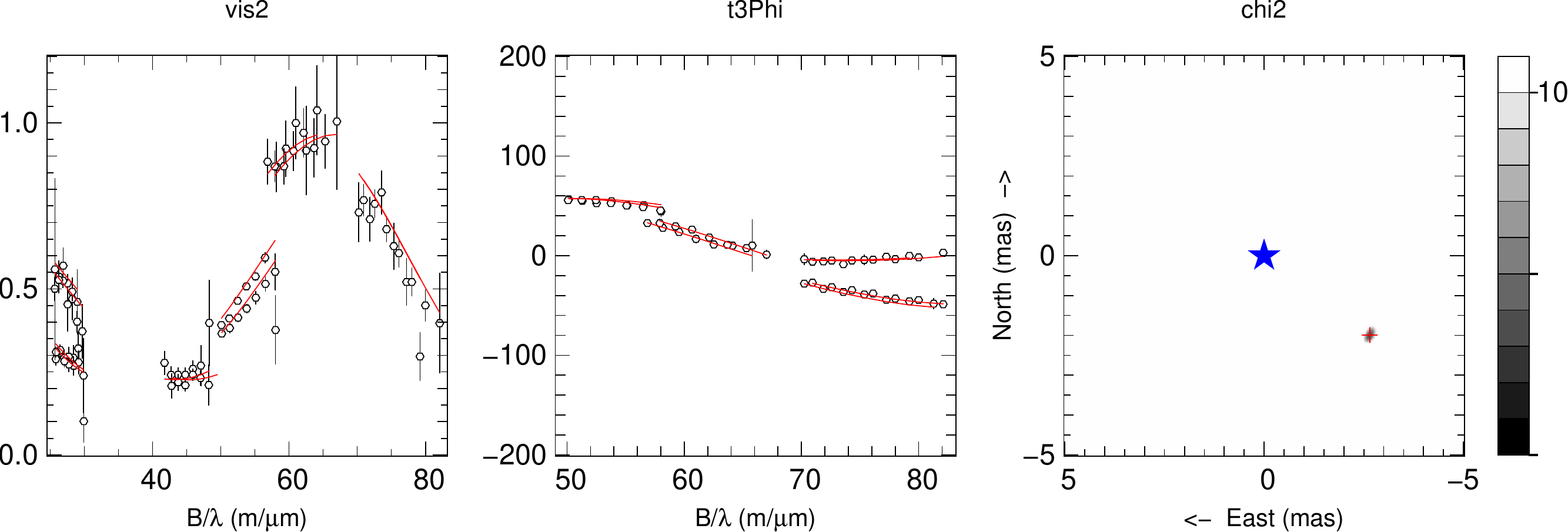}
\caption{Same as Fig.~\ref{fig:fit_interfdata_a} for the PIONIER observation of 2012-03-02.}\label{fig:fit_interfdata_b}\vspace{0.5cm}
\includegraphics[width=0.83\textwidth]{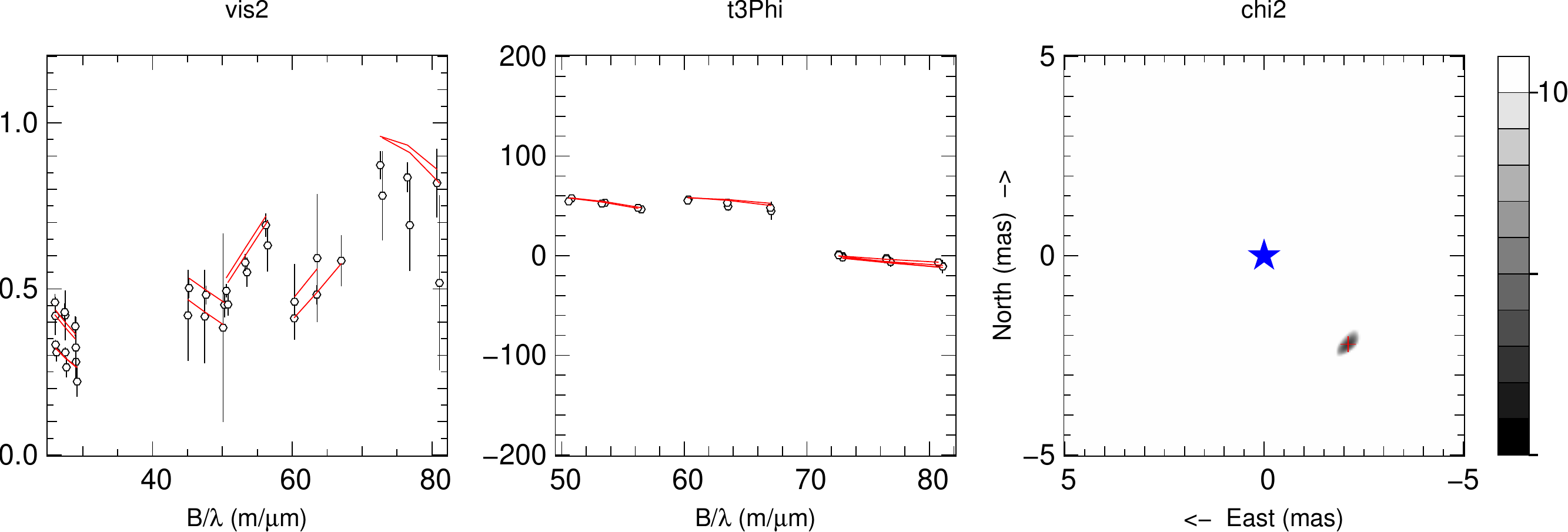}
\caption{Same as Fig.~\ref{fig:fit_interfdata_a} for the PIONIER observation of 2012-03-03.}\label{fig:fit_interfdata_c}\vspace{0.5cm}
\includegraphics[width=0.83\textwidth]{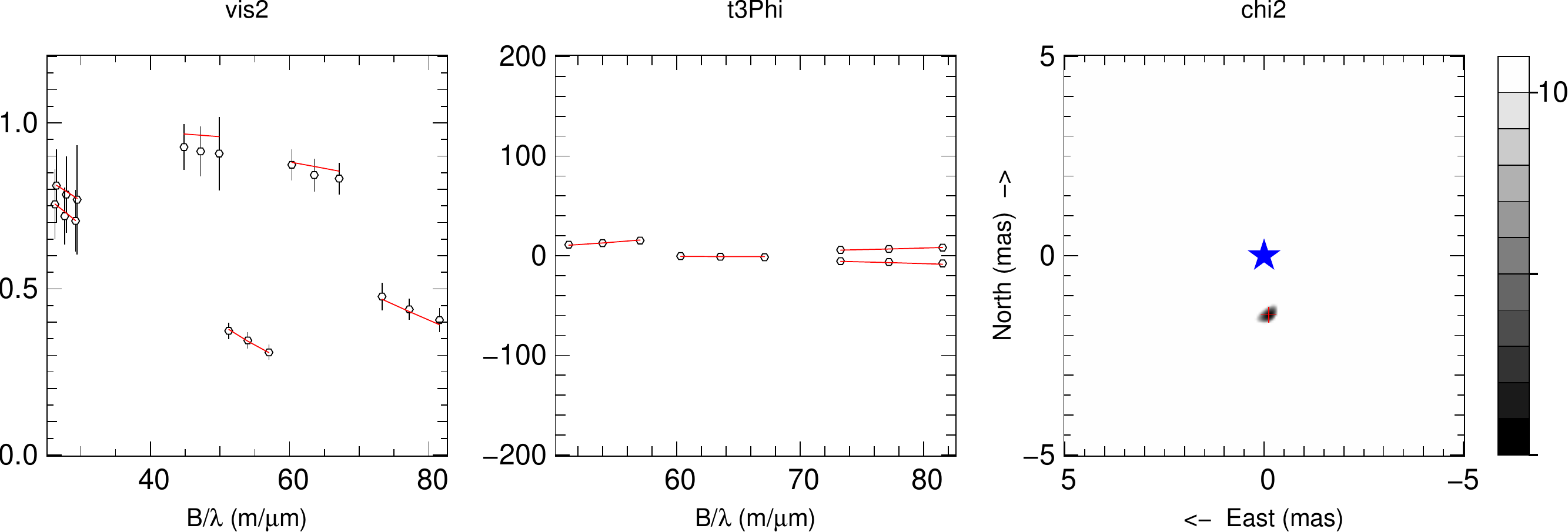}
\caption{Same as Fig.~\ref{fig:fit_interfdata_a} for the PIONIER observation of 2012-03-05.}\label{fig:fit_interfdata_d}
\end{figure*}

\section{Computation of characteristic times for circularization and synchronization}

\subsection{Equilibrium tide}

\citet{Press:1975} provide explicit formulas. For each component of the binary, we have
\begin{eqnarray}
t_\mathrm{circ} & = & \frac{125}{242}\frac{R_T}{K\delta N}(1-e^2)^5
\left(\frac{a}{R}\right)^{11}\frac{m_1^3}{m_2^2(m_1+m_2)}\qquad;\\
t_\mathrm{sync} & = & \frac{75}{224}\frac{R_Tg^2}{K\delta n}(1-e^2)^{9/2}
\left(\frac{a}{R}\right)^{9}\left(\frac{m_1}{m_2}\right)^3\qquad.
\end{eqnarray}
Here $N$ is the angular rotation velocity of the star, $g$ is its gyration 
radius (the moment of inertia is defined as $g^2mR^2$), $R_T$ is an effective 
Reynolds number that can be taken $\sim 20$, and $K$ and $\delta$ are numerical
parameters. According to \citet{Press:1975}, we can typically assume 
$K=0.025$ and $\delta=0.1$. The other symbols have 
their meaning introduced above. These formulas are valid for either 
component of the binary. In each case, $m_1$ stands for the mass of the 
star we are considering, and $m_2$ for the mass of the other star.

\subsection{Dynamical tides}

\citet{Zahn:1977} gave characteristic times.
\begin{eqnarray}
\frac{1}{t_\mathrm{sync}} & = & 
52^{2/3}\sqrt{\frac{Gm_1}{R^3}}\frac{m_2^2}{m_1^2}
\left(1+\frac{m_2}{m_1}\right)^{5/6}
\frac{E_2}{g^2}\left(\frac{R}{a}\right)^{17/2}\qquad;\\
\frac{1}{t_\mathrm{circ}} & = & \frac{21}{2}\sqrt{\frac{Gm_1}{R^3}}
\frac{m_2}{m_1}\left(1+\frac{m_2}{m_1}\right)^{11/6}
E_2\,\left(\frac{R}{a}\right)^{21/2}\qquad.
\end{eqnarray}
The conventions are the same as above. $E_2$ is a dimensionless constant characteristizing the strength of the dynamical tide. \citet{Zahn:1975} provided a table of $E_2$ and gyration radius at ZAMS for various masses. We interpolated these values for the actual masses and computed the characteristic times, assuming $n=N$ and $e=0$.

\subsection{Tidally driven meridional currents}

According to \citet{Tassoul:1992}, the characteristic times in years are
\begin{equation}
t_{\rm circ} =
\frac{14.4\times 10^{-N/4}}{g^2(1+m_2/m_1)^{11/8}}
\left(\frac{L_\odot}{L}\right)^{1/4}
\left(\frac{M_\odot}{m_1}\right)^{1/8}\left(\frac{R}{R_\odot}\right)^{9/8}
\left(\frac{a}{R}\right)^{49/8}
\end{equation}
and
\begin{equation}
t_{\rm sync} =
\frac{14.4\times 10^{-N/4}m_2}{m_1(1+m_2/m_1)^{3/8}}
\left(\frac{L_\odot}{L}\right)^{1/4}
\left(\frac{M_\odot}{m_1}\right)^{1/8}\left(\frac{R}{R_\odot}\right)^{9/8}
\left(\frac{a}{R}\right)^{33/8}.
\end{equation}

$L$ is the stellar luminosity and $N$ is a charateristic exponent that can be taken as 10 for stars with a convective envelope, and 0 for stars with a radiative envelope, which is the case here.

\end{document}